\documentclass[prl,twocolumn,groupedaddress,showpacs,floatfix]{revtex4}

\usepackage{graphicx}
\usepackage{bm}
\usepackage{amsmath}
\usepackage{amsfonts}
\usepackage{amssymb}
\usepackage{latexsym}

\newcommand{\ket}[1]{|#1\rangle}

\begin{document}

\title{Repeat-Until-Success Quantum Computing}
\author{Yuan Liang Lim,$^{1}$ Almut Beige,$^{2,1}$ and Leong Chuan Kwek$\,^{3,4}$}
\affiliation{
$^1$Blackett Laboratory, Imperial College London, Prince Consort Road, London SW7 2BW, United Kingdom \\
$^2$Department of Applied Mathematics and Theoretical Physics, University of Cambridge, Wilberforce Road, Cambridge CB3 0WA, United Kingdom \\
$^3$National Institute of Education, Nanyuang Technological University, Singapore 639 798, Singapore \\
$^4$Department of Physics, National University of Singapore, Singapore 117 542, Singapore}
\date{\today}

\begin{abstract}
We demonstrate the possibility to perform distributed quantum computing using {\em only} single photon sources (atom-cavity-like systems), linear optics and photon detectors. The qubits are encoded in stable ground states of the sources. To implement a universal two-qubit gate, two photons should be generated simultaneously and pass through a linear optics network, where a measurement is performed on them. Gate operations can be repeated until a success is heralded without destroying the qubits at any stage of the operation. In contrast to other schemes, this does not require explicit qubit-qubit interactions, a priori entangled ancillas nor the feeding of photons into photon sources.
\end{abstract}
\pacs{03.67.Lx, 42.50.Dv}

\maketitle

Generically, the construction of universal two-qubit gates for quantum computation can be classified under two main groups. The first group involves the realisation of entangling gates with the help of coherently controlled explicit qubit-qubit interactions. Examples of such schemes are universal quantum gates using nuclear magnetic resonance techniques \cite{chuang} and linear ion traps \cite{Schmidt-Kaler,Wineland}. In the second group, the qubit-qubit interactions are, at least partially, replaced by measurements. Universal two-qubit gates are obtained using ancillary qubits and by performing appropriate measurements after the system underwent a unitary time evolution. 

One of the most famous examples for measurement based quantum computing is the one-way quantum computer by Raussendorf and Briegel \cite{Raussendorf}, where a highly entangled cluster state is used as a resource. Another one is the Knill-Laflamme-Milburn (KLM) proposal for linear optics \cite{KLM}, where many photons pass through a linear optics network and measurements are performed on some of the output ports of the setup. With a certain probability, the remaining final state differs from the input only by the desired gate operation. However, if the measurement does not yield a certain result, the photons no longer contain the information initially stored in the qubits. In this case, the gate operation has failed and the whole computation has to be repeated. 

Other measurement based quantum computing schemes employ certain features of matter-photon interactions. They can provide, in principle, arbitrary high success rates, but are experimentally much more demanding. Examples are the cavity photon-assisted gates between atomic qubits based on an environment-induced quantum Zeno effect \cite{beige,pachos}, the ion-photon mapping scheme by Duan {\em et al.} \cite{Duan} and the universal photon gate using atom-doped optical fibres by Franson {\em et al.} \cite{Franson}. Simplifying experimental setups and increasing robustness and scalability of quantum computing architectures are among the many motivations for using measurements to process quantum information.

In this paper we propose an alternative use of measurements, which is especially advantageous for distributed quantum computing \cite{Grover,Cirac}. Distributed quantum computing aims at performing arbitrary computational tasks between the unknown quantum states of the distant nodes of a large network. In the following, we assume that these nodes are formed by sources for the generation of single photons on demand (atom-cavity like systems) \cite{Law,scheel}, which encode the logical qubits. We show that the implementation of the universal two-qubit phase gate
\begin{equation} \label{CZ}
U_{\rm CZ} = {\rm diag} \, ( \, 1, \, 1, \, 1, \, -1 \, ) 
\end{equation}
requires only the generation of two photons. These are then passed through a linear optics network and absorbed in a photon pair measurement process.  The proposed gate can be operated {\em repeatedly} until success without destroying the qubits at any stage of the computation. It is therefore possible to perform each gate operation in an eventually deterministic fashion, which is important for scalability. The generic final state is the desired output, if the measurement is deemed a success, or a new state, from which the original input state can be recovered easily by local operations. 

In contrast to previous measurement based schemes between distant photon sources \cite{Grangier,Zou}, we propose a scheme, where the gate success probability can in principle be as high as $100\%$. Moreover, we do not require local two-qubit gates between two trapped ions or atoms as it is the case in Refs.~\cite{Duan,Bose}. Compared to non-measurement based schemes for distributed quantum computing \cite{Mancini,VanEnk}, we do not require feeding of photons into a cavity. We also avoid the use of a priori entangled ancillas as a resource \cite{Eisert,Raussendorf}.

Repeat-Until-Success  quantum computing combines the advantages of quantum computing with {\em stationary} qubits (the single photon sources) and {\em flying} qubits (free-propagating single-photon qubits). On one hand, the considered setup provides good quantum memory due to the long decoherence times of the inner ground states of the stationary qubits. It is also relatively easy to implement single qubit rotations and to read out information with a very high precision. On the other hand, the use of flying qubits allows for a robust communication between arbitrary nodes of the network. At the same time, we avoid vulnerability to decoherence by avoiding the coherent control of explicit qubit-qubit interactions and the finite gate success rates of purely linear optics based quantum computing schemes. 

\begin{figure}
\begin{minipage}{\columnwidth}
\begin{center}
\resizebox{\columnwidth}{!}{\rotatebox{0}{\includegraphics{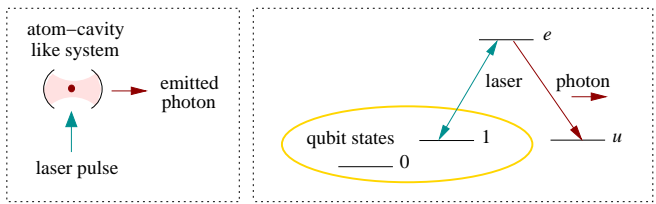}}}
\end{center}
\vspace*{-0.5cm} \caption{Schematic view of a single photon source and level configuration of the atomic structure containing the qubit.} \label{moon1}
\end{minipage}
\end{figure}

A possible level configuration of the required atom-cavity like single photon source is shown in Figure \ref{moon1}. The presence of one atom allows for the generation of one photon at the time while the cavity fixes its direction. The two atomic ground states $|0 \rangle$ and $|1\rangle $ encode one logical qubit. Photons should be generated such that an arbitrary single qubit state transforms according to
\begin{equation} \label{encoding}
|\psi_{\rm in} \rangle = \alpha \, |0 \rangle + \beta \, |1 \rangle ~~ \longrightarrow ~~ |\psi_{\rm enc} \rangle = \alpha \, |0;{\sf E} \rangle + \beta \, |1;{\sf L} \rangle \, .
\end{equation}
Here $|{\sf E} \rangle$ and $|{\sf L} \rangle$ denote the states of a single photon generated at an early and a late time, respectively. One way to implement the encoding (\ref{encoding}) is to first swap the atomic states $\ket{0}$ and $\ket{1}$. Then a laser pulse with increasing Rabi frequency should excite the $1$-$e$ transition (see Figure \ref{moon1}). This transfers the atom into the state $\ket{u}$ and places one excitation into the field of the strongly coupled optical cavity, if the atom was initially prepared in $|0 \rangle$. The photon then leaks out through the outcoupling mirror of the resonator \cite{Law}. The encoding operation (\ref{encoding}), which is feasible with present technology \cite{Kuhn2}, is completed by transfering $|u \rangle$ back into $|1 \rangle$, swapping again the states $|0 \rangle$ and $|1 \rangle$ and repeating the above described photon generation process. The final state $|\psi_{\rm enc} \rangle$ contains the initial qubit in the state of the source as well as in the state of the newly generated photon.  

Suppose, the two photon sources involved in a gate operation are initially prepared in the arbitrary state
\begin{equation} \label{in}
|\psi_{\rm in} \rangle = \alpha \, |00 \rangle + \beta \, |01 \rangle + \gamma \, |10 \rangle + \delta \, |11 \rangle \, .
\end{equation}
The state of the system then equals after performing the encoding (\ref{encoding}) for each photon source
\begin{equation} \label{enc2}
|\psi_{\rm enc} \rangle = \alpha \, |00;{\sf EE} \rangle + \beta \, |01;{\sf EL} \rangle + \gamma \, |10;{\sf LE} \rangle + \delta \, |11;{\sf LL} \rangle \, . 
\end{equation}
Since both qubits are redundantly encoded, a gate operation can be realised by performing a measurement on the photons. An important requirement for this to work is to choose the measurement basis such that none of the possible outcomes reveals any information about the coefficients $\alpha$, $\beta$, $\gamma$ and $\delta$. In fact, such bases exist and are known as {\em mutually unbiased bases} with respect to the computational basis formed by the states $|{\sf EE} \rangle$, $|{\sf EL} \rangle$, $|{\sf LE} \rangle$ and $|{\sf LL} \rangle$ \cite{Wooters}. Each mutually unbiased basis vector is of the form
\begin{equation} \label{meas}
|\Phi \rangle = {\textstyle {1 \over 2}} \big( |{\sf EE} \rangle + {\rm e}^{{\rm i} \varphi_1} \, |{\sf EL} \rangle +
{\rm e}^{{\rm i} \varphi_2} \, |{\sf LE} \rangle +  {\rm e}^{{\rm i} \varphi_3} \, |{\sf LL} \rangle \big) \, .
\end{equation}
Its detection combined with an absorption of the photon pair projects the state (\ref{enc2}) onto 
\begin{equation} \label{fin}
|\psi_{\rm fin} \rangle = \alpha \, |00 \rangle + \beta \, {\rm e}^{-{\rm i} \varphi_1} \, |01 \rangle + \gamma \,  {\rm
e}^{-{\rm i} \varphi_2} \, |10 \rangle  + \delta \,  {\rm e}^{-{\rm i} \varphi_3} \, |11 \rangle \, ,
\end{equation} 
which differs from the initial state (\ref{in}) only by the unitary operation 
\begin{equation}
U_{\rm phase}={\rm diag} \, \big( \, 1, \, {\rm e}^{-{\rm i} \varphi_1}, \, {\rm e}^{-{\rm i} \varphi_2}, \, {\rm e}^{-{\rm i} \varphi_3} \, \big) \, .
\end{equation}
Whenever $\varphi_3=\varphi_1+\varphi_2$, the measured outcome (\ref{meas}) is a product state and its detection imposes local operations onto the initial state (\ref{in}). However,  when $\varphi_3=\varphi_1+\varphi_2 + \pi$, the photons are detected in a maximally entangled state and the operation $U_{\rm phase}$ differs from the desired entangling gate (\ref{CZ}) only by local operations \cite{Grangier}.

Linear optics does not allow for complete Bell measurements with unit efficiency \cite{Lutkenhaus}. Instead we consider in the following a measurement basis, which distinguishes two maximally entangled states and two product states of the general form
\begin{eqnarray} \label{partialbell}
&& |\Phi_{1,2} \rangle \equiv {\textstyle {1 \over \sqrt{2}}} \, \big( |{\sf x}_1 {\sf y}_2 \rangle \pm |{\sf y}_1 {\sf x}_2 \rangle \big) \, , \nonumber \\
&& \ket{\Phi_3} \equiv |{\sf x}_1 {\sf x}_2 \rangle \, , ~~ \ket{\Phi_4} \equiv |{\sf y}_1 {\sf y}_2 \rangle \, .
\end{eqnarray} 
Here $|{\sf x}_i \rangle$ and $|{\sf y}_i \rangle$ are two orthogonal states describing a single photon produced by source $i$. One way to ensure that the basis states $|\Phi_i \rangle$ are mutually unbiased (see Eq.~(\ref{meas})) is to choose them such that 
\begin{eqnarray}\label{M1M2}
&& |{\sf x}_1 \rangle =  {\textstyle {1 \over \sqrt{2}}} \, \big( |{\sf E} \rangle + |{\sf L} \rangle \big) \, , ~~
|{\sf y}_1 \rangle =  {\textstyle {1 \over \sqrt{2}}} \, \big( |{\sf E} \rangle - |{\sf L} \rangle \big) \, , \nonumber \\
&& |{\sf x}_2 \rangle =  {\textstyle {1 \over \sqrt{2}}} \, \big( |{\sf E} \rangle + |{\sf L} \rangle \big) \, , ~~ |{\sf y}_2 \rangle  =  {\textstyle {1 \over \sqrt{2}}} \, {\rm i} \,  \big( |{\sf E} \rangle - |{\sf L} \rangle \big) \, , ~~~
\end{eqnarray}
since this definition implies
\begin{eqnarray} \label{Phi}
&& |\Phi_{1,2} \rangle = \pm {\textstyle {1 \over 2}} \, {\rm e}^{\pm{\rm i} \pi/4}  \, \big( |{\sf EE} \rangle \mp {\rm i} \, |{\sf EL} \rangle \pm {\rm i} \,  |{\sf LE} \rangle - |{\sf LL} \rangle \big) \, , ~~ \nonumber \\
&& |\Phi_3 \rangle =  {\textstyle {1 \over 2}} \, \big( |{\sf EE} \rangle + |{\sf EL} \rangle + |{\sf LE} \rangle + |{\sf LL} \rangle \big) \, , \nonumber \\
&& |\Phi_4 \rangle =  {\textstyle {1 \over 2}} \, {\rm i} \, \big( |{\sf EE} \rangle -  |{\sf EL} \rangle - |{\sf LE} \rangle + |{\sf LL} \rangle \big) \, .
\end{eqnarray}
If we express the photon states in this basis, the encoded two-qubit state (\ref{enc2}) becomes 
\begin{equation} \label{total}
\ket{\psi_{\rm enc}}= {\textstyle {1 \over 2}} \, \sum_{i=1}^4 \ket{\psi_i}\ket{\Phi_i} 
\end{equation}
with
\begin{eqnarray} \label{projection}
&& \ket{\psi_1} = {\rm e}^{-{\rm i} \pi/4} \, Z_1\big({\textstyle {1 \over 2}} \pi \big) \, Z_2 \big(-{\textstyle {1 \over 2}} \pi \big) \, U_{CZ} \, \ket{\psi_{\rm in}} \, , \nonumber \\
&& \ket{\psi_2} = -{\rm e}^{{\rm i} \pi/4} \, Z_1\big(-{\textstyle {1 \over 2}} \pi \big) \, Z_2 \big({\textstyle {1 \over 2}} \pi \big) \, U_{CZ} \, \ket{\psi_{\rm in}} \, , \nonumber \\ 
&& \ket{\psi_3} = \ket{\psi_{\rm in}} \, , ~~ \ket{\psi_4} = -  {\rm i} \,  Z_1(\pi) \, Z_2(\pi) \, \ket{\psi_{\rm in}} \, , 
\end{eqnarray}
where $Z_i(\varphi) = {\rm diag} \, (1, {\rm e}^{-{\rm i}\varphi})$ denotes a state dependent single qubit operation on atom $i$.

Eqs.~(\ref{total}) and (\ref{projection}) show that the controlled phase gate (\ref{CZ}) can indeed be implemented via a measurement of the states $\ket{\Phi_i}$. Each measurement outcome occurs with probability ${1 \over 4}$. The detection of the maximally entangled states $|\Phi_1 \rangle$ or $|\Phi_2 \rangle$ results in the implementation of the desired operation up to local phase gates. Finding the photon pairs in $|\Phi_3 \rangle$ and  $|\Phi_4 \rangle$ yields the initial state (\ref{in}) and the initial state (\ref{in})  up to a local conditional sign flip on both qubits, respectively. On average, the completion of the gate (\ref{CZ}) requires two repetitions of the above described process.

\begin{figure}
\begin{minipage}{\columnwidth}
\begin{center}
\resizebox{\columnwidth}{!}{\rotatebox{0}{\includegraphics{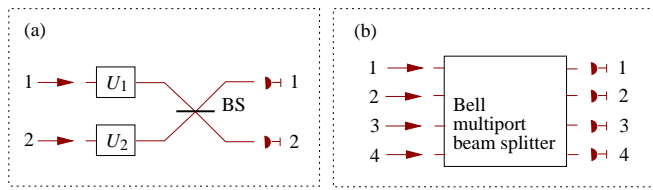}}}
\end{center}
\vspace*{-0.5cm} 
\caption{Linear optics networks for the realisation of a measurement of the basis states (\ref{Phi}) after encoding the photonic qubits in the polarisation degrees of two photons (a) or into four different spatial photon modes (b) involving either a beam splitter (BS) or a $4 \times 4$ Bell multiport beam splitter.} \label{moon2} 
\end{minipage}
\end{figure}

Finally, we describe two possible experimental realisations of the photon pair measurement (\ref{Phi}) with linear optics. One possibility is to translate the time bin encoding into a polarisation encoding, such that an early photon becomes a horizontally polarised one $(\ket{\sf E} \to \ket{\sf h})$ and a late photon becomes a vertically polarised one $(\ket{\sf L} \to \ket{\sf v})$. It is well known that sending two polarisation encoded photons through the different input ports of a 50:50 beam splitter together with polarisation sensitive measurements in the $\ket{\sf h}/\ket{\sf v}$-basis in the output ports would result in a measurement of the states ${\textstyle {1 \over {\sqrt 2}}}(\ket{\sf hv}\pm\ket{\sf vh})$, $\ket{\sf hh}$ and $\ket{\sf vv }$. To measure the states (\ref{partialbell}), we therefore propose to proceed as shown in Figure \ref{moon2}(a) and to perform the mapping $U_i = |{\sf h} \rangle \langle {\sf x_i}| + |{\sf v} \rangle \langle {\sf y}_i|$ on the photon coming from source $i$. Using Eq.~(\ref{M1M2}), we see that this corresponds to the single qubit rotations 
\begin{eqnarray}
U_1 &=&  {\textstyle {1 \over \sqrt{2}}} \, \big[ \, |{\sf h} \rangle \big( \langle {\sf h}| + \langle {\sf v}| \big) +  |{\sf v} \rangle \big( \langle {\sf h}| - \langle {\sf v}| \big) \, \big] \, , \nonumber \\
U_2 &=&  {\textstyle {1 \over \sqrt{2}}} \, \big[ \, |{\sf h} \rangle \big( \langle {\sf h}| + \langle {\sf v}| \big) - {\rm i} \, |{\sf v} \rangle \big( \langle {\sf h}| -  \langle {\sf v}| \big) \, \big] \, .
\end{eqnarray}
After leaving the beam splitter, the photons should be detected in the $\ket{\sf h}/\ket{\sf v}$-basis. Finding two photons of different polarisation in the same (different) detectors corresponds to a detection of $|\Phi_1 \rangle$ ($|\Phi_2 \rangle$). A detection of two ${\sf h}$ (${\sf v}$) polarised photons indicates a measurement of $|\Phi_3 \rangle$ ($|\Phi_4 \rangle$). 

Alternatively, one can redirect the generated photons to the different input ports of a $4 \times 4$ Bell multiport beam splitter as shown in Figure \ref{moon2}(b). If $a_n^\dagger$ ($b_n^\dagger$) denotes the creation operator for a photon in input (output) port $n$, the effect of the multiport can be summarised as \cite{multi}
\begin{equation} \label{scatter}
a_n^\dagger \to \sum_m U_{mn} b_m^\dagger ~~ {\rm with} ~~ 
U_{nm} = {\textstyle {1 \over 2}} \, {\rm i}^{(n-1)(m-1)} \, .
\end{equation}
A Bell multiport redirects each incoming photon with equal probability to any of the possible output ports and can therefore be used to erase the which-way information of the incoming photons. One way to measure in the mutually unbiased basis (\ref{Phi}) is to direct an early (late) photon from source 1 to input port 1 (3) and to direct an early (late) photon from source 2 to input port 2 (4). If $\ket{\rm vac}$ denotes the state with no photons in the setup, this results in the conversion $|{\sf EE} \rangle \to a_1^\dagger a_2^\dagger \, \ket{{\rm vac}}$, $|{\sf EL} \rangle \to a_1^\dagger a_4^\dagger \, \ket{{\rm vac}}$, $|{\sf LE} \rangle \to a_2^\dagger a_3^\dagger \, \ket{{\rm vac}}$ and $|{\sf LL} \rangle \to a_3^\dagger a_4^\dagger \, \ket{{\rm vac}}$. This conversion should be realised such that the photons enter the multiport at the same time. Using Eq.~(\ref{scatter}) one can show that the network transfers the basis states (\ref{Phi}) as
\begin{eqnarray} \label{multiportresult}
|\Phi_1 \rangle & \to & {\textstyle {1 \over {\sqrt 2}}} \, \big(  b_1^\dagger b_4^\dagger - b_2^\dagger 
b_3^\dagger \big) \, \ket{\rm vac} \, , \nonumber  \\ 
|\Phi_2 \rangle & \to & - {\textstyle {1 \over {\sqrt 2}}} \, \big(  b_1^\dagger b_2^\dagger - b_3^\dagger 
b_4^\dagger  \big) \, \ket{\rm vac} \, , \nonumber  \\
|\Phi_3 \rangle & \to &  {\textstyle {1 \over 2}} \, \big( b_1^{\dagger \, 2}  - b_3^{\dagger \, 2} \big) \,  \ket{\rm vac} \, , 
\nonumber \\ 
|\Phi_4 \rangle & \to & - {\textstyle {1 \over 2}} \, \big( b_2^{\dagger \, 2}  - b_4^{\dagger \, 2} \big) \, \ket{\rm vac} \, .
\end{eqnarray}
Finally, detectors measure the presence of photons in each of the possible output ports. The detection of a photon in ports 1 and 4 or in 2 and 3 indicates a measurement of the state $\ket{\Phi_1}$, while a photon in the ports 1 and 2 or in 3 and 4 indicates the state $\ket{\Phi_2}$. The detection of two photons in the same output port, namely in 1 or 3 and in 2 or 4, corresponds to a measurement of the state $|\Phi_3 \rangle$ and $|\Phi_4 \rangle$, respectively.

{\em In summary}, we introduced the idea of Repeat-Until-Success quantum computing based on an unusual choice of measurement basis. As long as the logical qubits can be redundantly encoded onto auxiliary particles, like the states of newly generated photons, the ability to perform partial Bell measurements is sufficient to implement an eventually deterministic universal two-qubit gate. If the measurement does not immediately result in the desired gate operation, the initial qubits can be restored using only single qubit rotations. The whole process can therefore be repeated until it results finally in the completion of the quantum gate.

As a concrete example we described a scheme for distributed quantum computing, which allows the realisation of arbitrary computational tasks on the unknown quantum states of the distant nodes of a network. Each node consists of a single photon source like a trapped atom, an NV colour centre or a quantum dot placed inside an optical cavity, an optical  fibre or in front of a large lens. It can even consist of an atomic ensemble as in the experiment by Matsukevich and Kuzmich \cite{Kuzmich}. The basic idea of using mutually unbiased basis measurements for quantum computing, which we presented here, works for {\em any} type of encoding and a variety of systems. Here we chose to use time-bin encoded single photons, since their generation can be done with a relatively simple level structure (see Figure \ref{moon1}). Alternatively, one could also use polarisation encoding and the measurement shown in Figure {\ref{moon2}(a), which is interferometrically robust. It is also possible to use frequency encoding when converting afterwards, for example, to spatial encoding and measuring the photons with the setup shown Figure {\ref{moon2}(b)). 

We believe that the described setup is feasible with present technology, since {\em all} its basic components have already been demonstrated experimentally. For example, Legero {\em et al.} \cite{Legero} observed the Hong-Ou-Mandel effect for photons generated from an atom-cavity system, which is the basic mechanism of the above described partial Bell measurements. Blinov {\em et al.} \cite{blinov} observed entanglement between a trapped atom and a single photon of the type described in Eq.~(\ref{encoding}). Long decoherence time are achieved by storing the qubits in atomic ground states. This also has the advantage that single qubit rotations and the read out of computational results can be done with a very high precision using standard ion trap techniques \cite{Schmidt-Kaler,Wineland}. Moreover, we do not rely on the coherent control of explicit qubit-qubit interactions, which are replaced by the experimentally less demanding state dependent generation of single photons.  

Distinguishing the photon states $|\Phi_i \rangle$ does {\em not} require photon-number resolving detectors. However, it requires in principle unit detector efficiencies $\eta$. In the above described implementation, measurements are performed on the states of single photons, which can only be realised with $\eta < 1$. Repeat-Until-Success quantum computing with single photon sources with a high fidelity is nevertheless feasible, if the described setup is used for the generation of a cluster state for one-way quantum computing \cite{Raussendorf,barrett}. As long as $\eta$ is sufficiently large, a high fidelity $N$-qubit cluster state of the stationary qubits can be build in a time that scales almost linearly in $N$. Once built, the realisation of an algorithm requires only local measurements. Since Repeat-Until-Success quantum computing combines the advantages of stationary and flying qubits, it opens new perspectives for the implementation of quantum information processing. 

{\em Acknowledgements}.~We thank H. J. Briegel, A. Bro\-waeys, D. E. Browne, A. K. Ekert, P. Grangier and M. Jones for stimulating discussions. Y.~L.~L.~acknowledges funding from the DSO National Laboratories in Singapore. A.~B.~acknowledges support from the Royal Society and the GCHQ. This work was supported in part by the European Union and the EPSRC.

\noindent 
After completion of this work, we learnt of related work by S. Barrett and P. Kok \cite{barrett}, whom we thank for agreeing to submit our manuscripts simultaneously. 
\end{document}